\documentclass{ws-ijgmmp}
\usepackage[utf8]{inputenc}
\usepackage[T1]{fontenc}
\usepackage[colorlinks,hyperindex,plainpages=false]{hyperref}
\usepackage[compress]{cite}
\usepackage{amsmath}
\usepackage{amssymb}
\usepackage{amsfonts}
\usepackage{accents}

\DeclareMathAlphabet{\mathds}{U}{BOONDOX-ds}{m}{n}
\newcommand{\dd}{\mathrm{d}}

\newcommand{\wb}[1]{\accentset{\mathrm{w}}{#1}}

\begin{document}

\markboth{Manuel Hohmann}
{A geometric view on local Lorentz transformations in teleparallel gravity}

\catchline{}{}{}{}{}
%
%%%%%%%%%%%%%%%%%%%%%%%%%%%%%%%%%%%%%%%%%%%%%%%%%%%%%%%%%%%%%%%%%%%%

\title{A geometric view on local Lorentz transformations in teleparallel gravity}

\author{Manuel Hohmann}

\address{Laboratory of Theoretical Physics, Institute of Physics, University of Tartu, W. Ostwaldi 1, 50411 Tartu, Estonia\\
\email{manuel.hohmann@ut.ee}}

\maketitle

\begin{history}
\received{(Day Month Year)}
\revised{(Day Month Year)}
\end{history}

\begin{abstract}
Local Lorentz transformations play an important role in teleparallel gravity theories, in which a tetrad is conventionally employed as a fundamental field variable describing the gravitational field. It is commonly understood that modifications of general relativity in the teleparallel framework break a certain notion of local Lorentz invariance, which is present in the pure tetrad formulation of such theories, while another notion present in the covariant formulation is preserved. We illuminate these different notions from a geometric perspective, and distinguish them from what is commonly understood as breaking of local Lorentz invariance in the context of gravity phenomenology. Based on physical arguments, we present a geometric interpretation of the dynamical fields in teleparallel gravity, which unified and refines the conventional approaches.
\end{abstract}


\section{Introduction}\label{sec:intro}
During the last decade, teleparallel gravity theories, which attribute the gravitational interaction to the torsion of a flat, metric-compatible connection instead of the curvature of the Levi-Civita connection, have received growing attention~\cite{Einstein:1928,Aldrovandi:2013wha}, and provided numerous models to address the open questions in gravity and cosmology~\cite{Bahamonde:2021gfp,Cai:2015emx,CANTATA:2021ktz}. An important notion, which is of particular relevance in these theories, is that of local Lorentz invariance, and its potential breaking in modified teleparallel gravity theories~\cite{Li:2010cg,Sotiriou:2010mv,Golovnev:2021omn}. By non-experts in the field, this is frequently perceived as a pathology of these theories, contradicting the observed local Lorentz invariance in nature. But also within the community of experts in teleparallel gravity it occasionally leads to debates.

The central origin of these debates is the fact that in its most conventional formulations, teleparallel gravity theories employ a tetrad as one of the fundamental field variables instead of a metric, possibly supplemented with a flat Lorentz spin connection in its covariant formulation. While in the teleparallel equivalent of general relativity (TEGR)~\cite{Maluf:2013gaa} only the components of the metric enter into the gravitational field equations, leaving any additional field components contained in the tetrad or spin connection as both undetermined and irrelevant gauge degrees of freedom for the description of the gravitational field, this is not the case in modified teleparallel gravity theories. The latter usually involve all tetrad components (in the non-covariant pure tetrad formulation) or combinations of the tetrad and spin connection components (in the covariant formulation), up to possibly remnant symmetries~\cite{Ferraro:2014owa,Chen:2014qtl,Maluf:2018coz,Guzman:2019ozl,Bejarano:2019fii,Ferraro:2020tqk}. This property is often called ``breaking of local Lorentz invariance in modified gravity'', as it breaks the freedom to perform local Lorentz transformations on the tetrad only, which does not affect the field equations of TEGR, but does affect the field equations of modified theories, in the sense that it is not a symmetry transformation on the space of solutions to these theories.

In the context of observational tests and the phenomenology of gravitational theories, the term local Lorentz invariance is conventionally used synonymously with the absence of preferred frame effects, in the sense that the outcome of a local, freely falling, non-gravitational experiment should not depend on the velocity of the reference frame~\cite{Will:2018bme,Will:2014kxa}. Breaking this symmetry would lead to observational consequences, which have not been observed in nature, putting strict bounds on any possible violation of local Lorentz invariance in this sense. However, it should be emphasized that this concept is essentially different from the kind of local Lorentz invariance breaking encountered in teleparallel gravity mentioned above. The aim of this article is to illuminate these different notions of Lorentz invariance and its breaking from a geometric point of view, and provide a unified geometric picture relating both Lorentz invariance and its breaking to the geometry of spacetime as perceived from an observer's perspective.

This article is structured as follows. In section~\ref{sec:observer}, we motivate the use of tetrads from an observer's perspective, and discuss their role for performing observations of physical quantities, as well as their theoretical description. We then discuss their role as a fundamental field variable in teleparallel gravity in section~\ref{sec:tetrad}. In section~\ref{sec:palatini}, we discuss the alternative Palatini formulation of teleparallel gravity, which does not make use of tetrads. A unifying geometric interpretation of these different approaches and the fundamental teleparallel geometry is provided in section~\ref{sec:geometry}. We end with a conclusion in section~\ref{sec:conclusion}.

\section{The observer's perspective}\label{sec:observer}
We start our discussion from the viewpoint of an observer, whose intention is to perform measurements of physical quantities, and who establishes a local reference system for these measurements purely based on the observation of physical phenomena. In section~\ref{ssec:tetradobs}, we discuss how such observations can be used to determine a class of local frames, or tetrads, which provide such a local reference system. We then discuss the consequences of changing this reference system and the question of local Lorentz invariance in section~\ref{ssec:lorentz}.

\subsection{The role of tetrads in observations}\label{ssec:tetradobs}
Tetrads occur in different roles in relativistic physics. Here we start by motivating their notion from the perspective of an observer, who follows a worldline through spacetime \(M\), and carries along several types of measuring devices:
\begin{enumerate}
\item
The tangent vector to the worldline is used to define a future-pointing timelike vector at every point along the worldline, which can be chosen as the zero component \(e_0 = e_0{}^{\mu}\partial_{\mu}\) of a frame \(e_a{}^{\mu}\).
\item
Using the clock postulate and the notion of proper time along the worldline, the tangent vector \(e_0\) can be normalized.
\item
By observing the propagation of light, a set of lightlike directions can be established, from which spatial directions can be deduced; this light cone structure establishes a notion of causal propagation of information, by setting the speed of light as defining the maximal propagation velocity.
\item
The spatial directions can further be normalized by measuring the arrival time in a radar experiment~\cite{Pfeifer:2014yua}, and chosen orthogonal, e.g., by observing the linear polarization of light. This establishes the light cone as null directions.
\item
Finally, by performing experiments which violate parity invariance, such as observing radioactive decay, a spatial orientation can be selected. Together with the time orientation defined by the tangent vector of the observer's trajectory, this yields an oriented, time oriented frame.
\end{enumerate}
In summary, by performing these experiments, observers can establish a set of time-oriented, oriented, orthonormal frames at each point in spacetime \(M\). The set of these frames forms a subset of the general linear frame bundle \(\mathrm{GL}(M)\).

So far we have neither assumed nor deduced any particular structure for the set of orthonormal frames. Assuming only a clock postulate, which relates the proper time used in the normalization to a parametrization-invariant arc length of the observer's world line, as well as a causal structure defined by light cones, one ends up at a Finsler spacetime~\cite{Pfeifer:2011tk,Hohmann:2021zbt}. In the following, however, we will restrict ourselves to metric spacetimes, i.e., assume that there exists a Lorentzian metric \(g_{\mu\nu}\) on \(M\), so that the orthonormal frames are the solutions to the equation
\begin{equation}\label{eq:metframe}
g_{\mu\nu}e_a{}^{\mu}e_b{}^{\nu} = \eta_{ab}\,,
\end{equation}
where \(\eta_{ab} = \mathrm{diag}(-1, 1, 1, 1)\) is the Minkowski metric. Note that the metricity condition does not determine the orientation or time orientation of the frames; these must additionally be imposed. Assuming that the spacetime \(M\) equipped with the metric \(g_{\mu\nu}\) is orientable and time-orientable, the orthonormal frame bundle \(\mathrm{O}(M, g)\) obtained as solutions to the metricity condition~\eqref{eq:metframe} consists of four connected components. Selecting an orientation and time orientation corresponds to selecting one of these components. We will denote this component by \(\mathrm{SO}_0(M, g, \epsilon)\), where \(\epsilon\) represents the chosen orientation and time orientation.

Finally, we remark that \(\mathrm{GL}(M)\), \(\mathrm{O}(M, g)\) and \(\mathrm{SO}_0(M, g, \epsilon)\) are principal bundles with structure groups \(\mathrm{GL}(4)\), \(\mathrm{O}(1,3)\) and \(\mathrm{SO}_0(1,3)\), respectively, where the latter denotes the connected component of the unit element in \(\mathrm{O}(1,3)\). The operation relating these bundles is known as reduction of the structure group~\cite{Baum:2014}; hence, we conclude that in a metric, oriented, time-oriented spacetime observers can establish a set of distinguished frames as a \(\mathrm{SO}_0(1,3)\) reduction of the general linear frame bundle. In other words, any two observer frames at a point \(x \in M\) which are compatible with the same metric, oriented, time-oriented spacetime are related by a proper Lorentz transformation \(\Lambda \in \mathrm{SO}_0(1,3)\).

\subsection{Lorentz covariance, invariance and preferred frame effects}\label{ssec:lorentz}
In the following we will assume that observers can perform measurements of physical quantities at every point \(x \in M\) which belongs to their worldline. Let \(Q, Q'\) denote the measurement outcomes obtained by two observers passing through the same point \(x\), but using different frames \(e, e'\) at \(x\), which are orthonormal with respect to a common metric and sharing the same orientation and time orientation. Hence, they are related by a proper Lorentz transformation \(\Lambda \in \mathrm{SO}_0(1,3)\) via
\begin{equation}
e'_a{}^{\mu} = e_b{}^{\mu}\Lambda^b{}_a\,.
\end{equation}
Physical observations show that the quantities we encounter in nature are \emph{Lorentz covariant}: the measurement outcomes \(Q, Q'\) mentioned above are related to each other by some representation \(\rho: \mathrm{SO}_0(1,3) \to \mathrm{GL}(n)\) of the Lorentz group, \(Q' = \rho(\Lambda)Q\), where \(n\) denotes the dimension of the representation. This means that \(Q, Q'\) can be regarded as different coordinate representations of a common object: this object is an element of the fiber over \(x\) of the associated fiber bundle \(\mathrm{SO}_0(M, g, \epsilon) \times_{\rho} \mathbb{R}^n\). Usually, \(\rho\) turns out to be a tensor representation, and hence this element is a tensor, whereas the bundle is a tensor bundle. Similarly, physical fields, which assign observables to every point \(x \in M\), will be sections of such tensor bundles, and hence tensor fields.

A related, yet different concept is that of \emph{local Lorentz invariance}: it states that the outcome of any local, non-gravitational test experiment is independent of the velocity of the (freely falling) apparatus~\cite{Will:2018bme,Will:2014kxa}. In other words, if we set up any non-gravitational experiment, such that the initial physical quantities have the values \(Q_0\) in a certain frame of reference, and the outcome is comprised of values \(Q_1\) in the same frame of reference, then by changing the initial values to \(Q_0' = \rho(\Lambda)Q_0\), one will obtain the outcome \(Q_1' = \rho(\Lambda)Q_1\). This means that the equations describing the physical laws which govern the evolution of the system must be Lorentz invariant, i.e., they must retain their form under local Lorentz transformations.

An alternative possibility to state the principle of local Lorentz invariance is as the absence of \emph{preferred frame effects}. By the latter we denote any observation which allows to distinguish a particular subclass of frames within the class of orthogonal frames. Such a distinction could be provided, for example, by a timelike background vector field, which determines a preferred time direction, and hence a preferred rest system. Note that, for example, a non-vanishing electromagnetic field in an experiment involving charged particles does \emph{not} establish a preferred frame in this sense, since under a local Lorentz transformation to a different frame \emph{all} physical quantities constituting the setup of the non-gravitational experiment, including the electromagnetic field, must be transformed, and both the Maxwell equations and the equations of motion of charged particles are invariant under this transformation. The principle of local Lorentz invariance rather demands the absence of any background fields which are independent of the experimental setup in a freely falling reference system.

One possibility to assess the local Lorentz invariance of a gravity theory is by using the parametrized post-Newtonian formalism~\cite{Will:2018bme,Will:2014kxa}. Here the breaking of local Lorentz invariance is signaled by the non-vanishing of certain parameters which characterize the theory under consideration. It is worth mentioning that in teleparallel gravity theories, in which ``local Lorentz invariance is broken'' in the sense we used this term in the introduction of this article, no preferred frame effects are present, and so no breaking of local Lorentz invariance in the sense we used in this section is observed~\cite{Ualikhanova:2019ygl,Emtsova:2019qsl,Flathmann:2019khc,Bahamonde:2020cfv}. This already shows that there is a fundamental difference between these notions, despite commonly denoting them by the same name, which we will explain further in the following sections.

\section{Tetrads and Lorentz transformations in teleparallel gravity}\label{sec:tetrad}
We now turn our focus to teleparallel gravity, and discuss the appearance of tetrads and local Lorentz transformations in the conventional formulations of these theories. We start with a brief review of the pure tetrad and covariant formulations of teleparallel gravity theories in section~\ref{ssec:tetforms}. We then clarify the relation between these formulations in section~\ref{ssec:weitz}, providing a constructive approach to obtain one from the other. Finally, we argue on the existence of such teleparallel geometries based on physical reasons in section~\ref{ssec:teleexist}.

\subsection{Pure tetrad and covariant formulations of teleparallel gravity}\label{ssec:tetforms}
Two different formulations of teleparallel gravity theories are conventionally used. In the original, pure tetrad formulation, the only fundamental variable describing the gravitational field is a tetrad, or coframe, which we will denote by \(\wb{\theta}^a{}_{\mu}\), for reasons which will become clear later. In mathematical terms, this is a section \(\wb{\theta}: M \to \mathrm{GL}^*(M)\) of the dual frame bundle. Its dual is a frame \(\wb{e}: M \to \mathrm{GL}(M)\), whose components are related to the coframe components by
\begin{equation}
\wb{e}_a{}^{\mu}\wb{\theta}^a{}_{\nu} = \delta^{\mu}_{\nu} \quad
\wb{e}_a{}^{\mu}\wb{\theta}^b{}_{\mu} = \delta_a^b\,.
\end{equation}
In the context of this formulation, local Lorentz transformations \(\Lambda: M \to \mathrm{SO}_0(1,3)\) are usually considered to act on the tetrad as
\begin{equation}\label{eq:tetradlortra}
\wb{\theta}^a{}_{\mu} = \Lambda^a{}_b\wb{\theta}'^b{}_{\mu}\,,
\end{equation}
and are \emph{not} symmetry transformations of the gravitational field equations, i.e., the field equations are not invariant under this transformation, and solutions are, in general, not mapped to solutions, except for TEGR~\cite{Ferraro:2011us,Tamanini:2012hg}. This is the reason for considering modified teleparallel gravity theories in this formulation as breaking local Lorentz invariance.

The aforementioned observation that the pure tetrad formulation of teleparallel gravity is not invariant under Lorentz transformations~\eqref{eq:tetradlortra} of its fundamental field variable has stipulated the development of the covariant formulation of teleparallel gravity~\cite{Krssak:2015oua,Golovnev:2017dox,Krssak:2018ywd,Hohmann:2018rwf}, in which next to the tetrad \(\theta^a{}_{\mu}\) a spin connection \(\omega^a{}_{b\mu}\) is employed as dynamical field variable. This spin connection is assumed to be antisymmetric, \(\omega^{(ab)}{}_{\mu} = 0\), and flat,
\begin{equation}
\partial_{\mu}\omega^a{}_{b\nu} - \partial_{\nu}\omega^a{}_{b\mu} + \omega^a{}_{c\mu}\omega^c{}_{b\nu} - \omega^a{}_{c\nu}\omega^c{}_{b\mu} = 0\,.
\end{equation}
In this formulation, local Lorentz transformations are considered to act jointly on both field variables as
\begin{equation}\label{eq:covarlortra}
\theta^a{}_{\mu} = \Lambda^a{}_b\theta'^b{}_{\mu}\,, \quad
\omega^a{}_{b\mu} = \Lambda^a{}_c(\Lambda^{-1})^d{}_b\omega'^c{}_{d\mu} + \Lambda^a{}_c\partial_{\mu}(\Lambda^{-1})^c{}_b\,,
\end{equation}
and are symmetry transformations of the gravitational field equations, i.e., the field equations are invariant under this transformation. In this sense, local Lorentz invariance is restored in the covariant formulation of teleparallel gravity.

Arguments have been presented in favor of either formulation. We will not repeat these arguments here, but rather focus on the common properties of both formulations.

\subsection{The Weitzenböck gauge}\label{ssec:weitz}
We start our discussion in this section from the perspective of the covariant formulation of teleparallel gravity. It is conventionally concluded from the antisymmetry and flatness of the spin connection that there exists a particular local Lorentz transformation \(\Lambda^a{}_b\) under which the spin connection transforms to vanish. As a consequence, one sometimes finds the spin connection expressed in terms of this Lorentz transformation as
\begin{equation}
\omega^a{}_{b\mu} = \Lambda^a{}_c\partial_{\mu}(\Lambda^{-1})^c{}_b\,,
\end{equation}
obtained from setting \(\omega'^a{}_{b\mu} = \wb{\omega}^a{}_{b\mu} = 0\) in the transformation~\eqref{eq:covarlortra}. The Lorentz gauge which is obtained by this transformation is then called the Weitzenböck gauge, and the tetrad can accordingly be written as
\begin{equation}\label{eq:wbtetrad}
\theta^a{}_{\mu} = \Lambda^a{}_b\wb{\theta}^b{}_{\mu}
\end{equation}
in terms of the tetrad \(\theta'^a{}_{\mu} = \wb{\theta}^a{}_{\mu}\) in the Weitzenböck gauge. Note that despite calling it \emph{the} Weitzenböck gauge, it is actually a family of gauges, which are related by a global Lorentz transformation
\begin{equation}
\Lambda^a{}_b \mapsto \Lambda'^a{}_b = \Lambda^a{}_c\Omega^c{}_b\,, \quad
\wb{\theta}^a{}_{\mu} \mapsto \wb{\theta}'^a{}_{\mu} = (\Omega^{-1})^a{}_b\wb{\theta}^b{}_{\mu}\,.
\end{equation}
with \(\partial_{\mu}\Omega^a{}_b = 0\). Further, it is often stated that by imposing the Weitzenböck gauge, the covariant formulation of teleparallel gravity reduces to the non-covariant formulation, in which the tetrad is the only dynamical field variable, and that these two formulations are therefore dynamically and phenomenologically equivalent. Instead of taking these statements for granted, we raise the question whether the aforementioned conclusion is indeed valid. To answer this question, we follow a constructive approach and explicitly derive the Lorentz transformation \(\Lambda^a{}_b\) to the Weitzenböck gauge.

In order to construct the Lorentz transformation \(\Lambda^a{}_b\) for a given teleparallel geometry, defined by a tetrad and a flat, antisymmetric spin connection, recall first that these quantities define two quantities which are invariant under local Lorentz transformations. These are a metric
\begin{equation}\label{eq:metric}
g_{\mu\nu} = \eta_{ab}\theta^a{}_{\mu}\theta^b{}_{\nu}
\end{equation}
and a flat, metric-compatible, affine connection as the unique solution of the ``tetrad postulate''
\begin{equation}\label{eq:tetpost}
\partial_{\mu}\theta^a{}_{\nu} + \omega^a{}_{b\mu}\theta^b{}_{\nu} - \Gamma^{\rho}{}_{\nu\mu}\theta^a{}_{\rho} = 0\,.
\end{equation}
The latter formula is Lorentz covariant, and hence retains its form under local Lorentz transformations, up to an overall factor \(\Lambda^a{}_b\); due to its free Lorentz index it thus transforms as a vector. It follows that it holds in particular also in the Weitzenböck gauge, where it reads
\begin{equation}\label{eq:tetpostwb}
\partial_{\mu}\wb{\theta}^a{}_{\nu} - \Gamma^{\rho}{}_{\nu\mu}\wb{\theta}^a{}_{\rho} = 0\,.
\end{equation}
One therefore finds that for every fixed value of the Lorentz index, the components of the Weitzenböck tetrad \(\wb{\theta}^a{}_{\mu}\) constitute a covariantly constant covector field. A solution to this equation, which reproduces both the metric and the affine connection, can be found by the following procedure:
\begin{enumerate}
\item
At some point \(x \in M\), choose a tetrad \(\wb{\theta}^a{}_{\mu}(x)\), such that it induces the metric~\eqref{eq:metric}.
\item
For any other point \(y \in M\), choose a path from \(x\) to \(y\), and use the covariant constancy~\eqref{eq:tetpostwb} to obtain \(\wb{\theta}^a{}_{\mu}(y)\) by parallel transport of \(\wb{\theta}^a{}_{\mu}(x)\).
\end{enumerate}
The metric compatibility of the tetrad obtained in the second step is guaranteed by the fact that the affine connection is metric compatible.

One notices two choices to be made in the construction above, namely the tetrad at the original point \(x\) and the paths towards the remaining points \(y\). Naturally the question arises how these choices influence the obtained solution. For the former, two different choices of the initial tetrad, which induce the same metric, are related by a Lorentz transformation, and so this freedom of choice corresponds to the residual gauge freedom under global Lorentz transformations, which preserves the Weitzenböck gauge condition of vanishing spin connection. For the latter, it is known that the parallel transport of a (co-)vector from one point to another in general depends on the path taken between these points. The teleparallel affine connection, however, is flat, and so one naively concludes that it defines a parallel transport which is independent of the chosen path. Nevertheless, one must keep in mind that this conclusion holds only for homotopic paths, and that a (spacetime) manifold, on which any two paths with the same endpoints are homotopic, is necessary simply connected. While this condition is satisfied for many examples considered in physics, it does not hold in general, and is violated, e.g., if one considers intra-spacetime wormholes.

We therefore conclude that in general the Weitzenböck gauge can be achieved only locally, but not globally. Hence, the covariant and pure tetrad formulations of teleparallel gravity can be regarded equivalent only for simply connected spacetimes.

\subsection{The existence of teleparallel geometries}\label{ssec:teleexist}
For our construction of the Weitzenböck gauge in the previous section we have assumed that our spacetime is already equipped with a global tetrad \(\theta^a{}_{\mu}\) and a flat spin connection \(\omega^a{}_{b\mu}\). However, this assumption contains an implicit restriction of the possible topologies of spacetime, since any manifold which admits a global section of the frame bundle is, by definition, parallelizable. Of course one may simply raise this condition to a postulate, and consider only such spacetimes for which it is satisfied. However, from a more fundamental perspective, one should raise the question whether such a postulate excludes any physically relevant spacetime models, or whether there is any possibility to demand a parallelizable spacetime based on physical arguments.

A thorough treatment shows that indeed the latter is the case, and that a physical argument can be found in particle physics. This is due to the fact that the latter describes various constituents of matter, known as fermions, which possess half-integer spin, by spinor fields~\cite{Rajan:2016qiy}. Such fields necessarily transform under a representation of the spin group \(\mathrm{Spin}(1,3) \cong \mathrm{SL}(2, \mathbb{C})\), which is a double cover of the proper Lorentz group \(\mathrm{SO}_0(1,3)\). Hence, they must be described by sections of a bundle which is associated to a double cover of the orthonormal, oriented, time-oriented frame bundle; such a double cover is known as a spin structure, and a manifold which admits a spin structure is usually simply called ``spin''. For a four-dimensional spacetime, however, it turns out that being spin is exactly equivalent to being parallelizable~\cite{Geroch:1968zm}. We may therefore conclude that any spacetime, which allows for the description of fermions by spinor fields, is necessarily parallelizable, and thus admits global tetrads.

The question for the existence of a flat, antisymmetric spin connection is answered as a simple corollary. Obviously, one may always choose the Weitzenböck spin connection \(\wb{\omega}^a{}_{b\mu} = 0\). Together with a global tetrad, on whose existence we have concluded above, this yields a flat, metric-compatible affine connection through the tetrad postulate~\eqref{eq:tetpostwb}. In the case of a simply connected parallelizable spacetime, any flat, metric-compatible affine connection is of this form, as it allows to reconstruct the global tetrad up to a global Lorentz transformation, as argued in the previous section. However, as we also argued, for a non-simply connected spacetime, more general flat connections may exist, which cannot be represented by a tetrad in the Weitzenböck gauge.

\section{The Palatini approach}\label{sec:palatini}
Another common description besides the usual tetrad formalism of teleparallel geometry, which attempts to avoid the gauge ambiguity in the covariant formulation, is the Palatini approach. We discuss this approach and its relation to the tetrad formalism in section~\ref{ssec:orbit}, and comment on the role of the teleparallel affine connection in section~\ref{ssec:affcon}.

\subsection{Orbits and equivalence classes}\label{ssec:orbit}
The basic idea of the Palatini approach to teleparallel gravity is to consider the metric \(g_{\mu\nu}\) and the coefficients \(\Gamma^{\mu}{}_{\nu\rho}\) of a flat, metric-compatible affine connection as fundamental fields instead of the tetrad and the spin connection~\cite{BeltranJimenez:2018vdo}. Here the vanishing curvature and nonmetricity of the affine connection is conventionally imposed via Lagrange multipliers, but also other methods are possible~\cite{Hohmann:2021fpr}. This approach has the advantage that both the metric and the affine connection, when defined from a teleparallel geometry in terms of a tetrad and spin connection using the relations~\eqref{eq:metric} and~\eqref{eq:tetpost}, are invariant under local Lorentz transformations of the form~\eqref{eq:covarlortra}, so that no gauge choice is necessary when solving the field equations of a teleparallel gravity theory. However, care must be taken that these are actually invariant under the \emph{full} Lorentz group \(\mathrm{O}(1,3)\), which also includes temporal and spatial reflection, and so they do not retain the information on the orientation and time orientation of spacetime, which is a necessary ingredient in the complete description of physical phenomena. Hence, the Palatini approach can be regarded as equivalent to the tetrad approach only of these orientations, for which we introduced the collective notation \(\epsilon\), are fixed.

From a geometric point of view, local Lorentz transformation form a group under pointwise multiplication, which acts on the space of pairs \((\theta^a{}_{\mu}, \omega^a{}_{b\mu})\) of tetrads and spin connections by the prescription~\eqref{eq:covarlortra}. In the covariant formulation of teleparallel gravity, this is a gauge transformation, and so any two teleparallel geometries belonging to the same orbit can be regarded as equivalent. Finally, there is a one-to-one correspondence between such orbits and tuples \((g_{\mu\nu}, \epsilon, \Gamma^{\mu}{}_{\nu\rho})\) of metrics, orientations, time orientations and teleparallel affine connections. In other words, these variables are simply an alternative parametrization of the space of orbits, which represents the space of inequivalent teleparallel geometries.

The main insight we shall take from the Palatini approach is that any teleparallel geometry, viewed as an equivalence class of pairs of tetrads and spin connections under (proper) local Lorentz transformations is equivalently described by a metric, orientation, time orientation and flat, metric-compatible affine connection. Note that none of these quantities determines a preferred frame at any point in spacetime, or impose any restriction on the possible frames an observer may choose in order to perform measurements or how these measurements are related between different frames. Any frames which are related by a proper Lorentz transformation are compatible with the same metric, orientation and time orientation, and thus yield an equivalent description of physical phenomena. No breaking of local Lorentz invariance is observed. To further clarify this point, we elaborate on the role of the teleparallel affine connection below.

\subsection{The role of the affine connection}\label{ssec:affcon}
The preceding discussion raises the question for the role of the teleparallel affine connection and its physical interpretation. We emphasize again that it does not single out a preferred frame at any point in spacetime; however, it does provide a parallel transport of tangent vectors, and hence frames, which is invariant under a homotopic change of the path. This fact is the origin of the term ``teleparallelism'', or ``Fernparallelismus''~\cite{Einstein:1928}. In this sense it does single out a system of tetrads in a simply connected neighborhood, which is obtained by parallel transport; this is simply the (local) Weitzenböck tetrad~\eqref{eq:wbtetrad}. However, keep in mind that this is defined only locally and up to a global Lorentz transformation, and so we will not regard it as a fundamental physical field.

In the following, we will assume the metric matter coupling prescription, according to which matter couples to gravity only via the metric, its Levi-Civita connection, and respects the orientation and time orientation, while there is no direct coupling to the teleparallel affine connection. It has been argued that this is the only consistent matter coupling in teleparallel gravity~\cite{Obukhov:2002tm,Maluf:2003fs,Mosna:2003rx,Mielke:2004gg,Obukhov:2004hv,BeltranJimenez:2020sih}. In this case the teleparallel affine connection makes its appearance only via its contribution to the gravitational action, where it determines the dynamics of the gravitational interaction. In this regard its role is not different from, e.g., the role of a scalar field describing dark energy in a scalar-tensor theory of gravity, or a second metric in a bimetric theory, which does not directly couple to matter.

With this understanding, we can now return to the question of broken local Lorentz invariance, which we encountered in different contexts. From the perspective of the Palatini formulation, Lorentz transformations of the form~\eqref{eq:tetradlortra} are simply transformations of the teleparallel affine connection \(\Gamma^{\mu}{}_{\nu\rho}\). Since the dynamics of the gravitational field in modified teleparallel gravity theories depends on this connection, it is clear that such transformations also change the gravitational dynamics, and lead to inequivalent geometries. However, what appears as inequivalent tetrads in the pure tetrad formulation of teleparallel gravity, simply appears as inequivalent teleparallel affine connection in the Palatini formulation. In TEGR, which is invariant under such transformations, the teleparallel affine connection fully decouples from the field equations, and so the aforementioned transformations become an internal symmetry of the theory, which is (at least partially) broken in modified theories. In particular it should be emphasized that this purely internal symmetry breaking does not constitute a breaking of local Lorentz invariance in the sense of preferred frame effects as discussed in section~\ref{ssec:lorentz}: observers remain free to use any orthonormal, oriented, time oriented frame defined by the metric~\eqref{eq:metric} and the orientations \(\epsilon\) of the spacetime geometry. The coframe \(\wb{\theta}^a{}_{\mu}\) is not a preferred coframe in the sense of observations, but simply a convenient possibility to combine the ``dark'' components of the affine connection, which couples only gravitationally, with the metric components.

Nevertheless, the aforementioned statements do not mean that the introduction of the teleparallel affine connection as additional field mediating the gravitational interaction may not lead to pathologies~\cite{Golovnev:2020zpv}. Possible issues may become manifest, for example, in the form of strong coupling issues, or the breakdown of perturbation theory around particular backgrounds~\cite{BeltranJimenez:2019nns,Jimenez:2020ofm,Golovnev:2020nln}. In order to avoid such problems, all physical degrees of freedom, i.e., those which directly or indirectly couple to physical observables, must obey a causally determined evolution~\cite{Kopczynski:1982,Nester:1988,Ong:2013qja,Izumi:2013dca,Golovnev:2018wbh,Golovnev:2020aon}. This does not exclude the possibility of true ``remnant'' gauge symmetries in the case that the aforementioned internal Lorentz symmetry is only partially broken, which would imply that certain components of the affine connection are neither determined by the field equations, nor relevant for the evolution of the observable fields~\cite{Ferraro:2014owa,Chen:2014qtl,Maluf:2018coz,Guzman:2019ozl,Bejarano:2019fii,Ferraro:2020tqk}.

\section{A unified geometric picture}\label{sec:geometry}
The discussions detailed in the previous sections raise the questions which geometric objects are most suitable as dynamical field variables in teleparallel gravity. We will now conclude on these discussions and attempt to answer this question constructively. In section~\ref{ssec:dynfield}, we summarize the different choices of the dynamical field variables and their relations. These are interpreted in section~\ref{ssec:unification} as different representatives of a common geometric structure. Finally, in section~\ref{ssec:spinor} we discuss the relation of the tetrad and the spin structure, in order to obtain a more concise notion of teleparallel geometry.

\subsection{Choices of the dynamical field variables}\label{ssec:dynfield}
We start from the viewpoint of the pure tetrad formulation of teleparallel gravity, in which the only gravitational field variable is a tetrad \(\wb{\theta}^a{}_{\mu}\), which gives rise to the Weitzenböck connection, whose torsion enters into the gravitational field equations. Clearly, a globally chosen tetrad allows to equip the spacetime manifold with all necessary quantities to describe observations as argued in section~\ref{ssec:tetradobs}: a metric, as well as a spatial and temporal orientation. Further, its existence guarantees the existence of a spin structure, as argued in section~\ref{ssec:teleexist}. However, it leaves the question whether the particular choice of the tetrad has any physical significance besides its gravitational coupling, since any other tetrad, which is related by a local Lorentz transformation, yields the same geometry probed by observables according to section~\ref{ssec:lorentz}, and would thus be indistinguishable by observations. The local Lorentz invariance inherent in observations is therefore not reflected by this choice of the dynamical variable and hence appears broken.

The aforementioned argument is a possible motivation for the covariant formulation of teleparallel gravity, where a flat Lorentz spin connection is considered as a dynamical, gravitational field variable next to the tetrad. In this case, any tetrad related by a local (and proper, i.e., orientation-preserving) Lorentz transformation is regarded as equivalent, and the spin connection must be transformed accordingly. The appealing consequence of this choice of the dynamical field variables is the natural appearance of Lorentz transformations relating physically equivalent tetrads. It is often claimed to be equivalent to the pure tetrad formulation, and that the latter is recovered by passing to the Weitzenböck gauge; however, as argued in section~\ref{ssec:weitz}, this is in general only possible locally, but not globally. Hence, we conclude that the covariant formulation of teleparallel gravity allows for more general geometries than the pure tetrad formulation, and is therefore favored. Nevertheless, this generality comes at the price of introducing unphysical gauge degrees of freedom, and so the question arises whether these can be avoided.

Following the discussion in section~\ref{ssec:orbit}, the Palatini formulation of teleparallel gravity appears to answer the aforementioned question, and provide local Lorentz invariance without introducing any gauge degrees of freedom. However, note that these variables are invariant under the full Lorentz group \(\mathrm{O}(1,3)\), which does not preserve the orientation and time orientation encoded in the tetrad, and so this information is lost in passing to the Palatini formulation, unless one incorporates these into additional, discrete variables. Only then one obtains a formulation of teleparallel gravity which is equivalent to the previously mentioned one, and allows for a proper description of parity-violating interactions.

\subsection{Unification using geometry}\label{ssec:unification}
From the viewpoint of differential geometry, the different choices of the dynamical field variables in teleparallel gravity discussed in the previous section simply represent different possibilities to describe a common geometric structure. To see this, recall from section~\ref{ssec:tetradobs} that the metric, orientation and time orientation define a class of compatible frames, which form a principal bundle \(\pi: P \to M\) over the spacetime manifold \(M\), which we denoted as \(P = \mathrm{SO}_0(M, g, \epsilon)\). In mathematical terms, this is a \emph{reduction of the structure group} with respect to the canonical inclusion \(\mathrm{SO}_0(1,3) \hookrightarrow \mathrm{GL}(4)\). Further, we assumed \(M\) to be parallelizable as of section~\ref{ssec:teleexist}, in order to admit a spin structure. Finally, the flat, affine connection of a teleparallel spacetime defines a principal connection in the frame bundle. Since the connection is assumed to be metric-compatible, it can be reduced to a flat principal connection \(\Omega \in \Omega^1(P, \mathfrak{h})\), i.e., an equivariant one-form on \(P\) taking values in the Lie algebra \(\mathfrak{h} = \mathfrak{o}(1,3)\) and satisfying \(\dd^{\Omega}\Omega = 0\). In the following, we will see how this most geometric picture relates to the common description of teleparallel gravity in terms of the previously mentioned field variables, and allows to recover these.

Since \(H = \mathrm{SO}_0(1,3)\) is a closed subgroup of \(G = \mathrm{GL}(4)\), there is an equivalent and more accessible description of the geometric setting given above~\cite{Baum:2014}. First note that there is a one-to-one correspondence between reductions of the structure group with respect to the canonical embedding and sections of the coset bundle \(\chi: C \to M\) with \(C = \mathrm{GL}(M) \times_{\lambda} G/H\), where \(\lambda: G \times G/H \to G/H\) denotes the canonical left action of \(G\) on the coset space \(G/H\). The latter has two connected components, corresponding to the two connected components of \(G\) given by the sign of the determinant. Hence, every fiber of \(C\) has two connected components. If \(M\) is orientable, which we assume here, then also \(C\) itself has two connected components, i.e., for every continuous path \(\gamma: [0, 1] \to C\) with \(\chi(\gamma(0)) = \chi(\gamma(1)) = x\), \(\gamma(0)\) and \(\gamma(1)\) lie in the same connected component of \(\chi^{-1}(x)\). Each component of \(C\) determines an \emph{orientation} on \(M\). A Lorentzian metric \(g_{\mu\nu}\) on \(M\) determines four points in each fiber of \(C\), two in each connected component, which have a fixed relationship: the two points in the same fiber are related by multiplication with the matrix \(-\mathds{1}_4 \in G\), corresponding to a simultaneous temporal and spatial reflection, while the two points in the other connected component are obtained by multiplication with \(\mathrm{diag}(-1,1,1,1)\) and \(\mathrm{diag}(1,-1,-1,-1)\), respectively. Choosing an orientation selects one of the two connected components, hence leaving only two points in each fiber. If the metric is chosen such that these two points in each fiber form two continuous sections of \(C\) (and are thus disconnected), then the pair \((M, g)\) is \emph{time orientable}; choosing one of the two sections determines a time orientation. Conversely, every section of \(C\) determines a Lorentzian metric, an orientation and a time orientation on \(M\), which is exactly the data we required for the description of observations. Hence, instead of the rather abstract notation of a principal bundle reduction, one can equivalently work with the more familiar notion of a section of a fiber bundle over spacetime.

To complete the translation of the geometric picture of a teleparallel geometry given at the beginning of this section, let us finally remark that the principal connection \(\Omega\) on \(P\) is in one-to-one correspondence with a metric-compatible affine connection with coefficients \(\Gamma^{\mu}{}_{\nu\rho}\) on \(M\), and that the latter also constitutes a section of a bundle over \(M\), which can be obtained as a jet bundle. It does not come as a surprise that the flatness of the principal connection corresponds to the flatness of this affine connection and vice versa. Hence, from a geometric point of view, a teleparallel spacetime is nothing else than a manifold \(M\) equipped with a section of the bundle \(C\), turning it into an oriented, time-oriented Lorentzian spacetime, which additionally features a flat, metric-compatible, affine connection. From this point of view, one easily recovers the more familiar descriptions as follows:

\begin{enumerate}
\item
As argued above, a section of \(C\) corresponds to a Lorentzian metric, an orientation and a time orientation on \(M\), while the flat principal connection \(\Omega\) encodes a flat, metric-compatible affine connection. Hence, the geometric picture is simply an interpretation of the Palatini formulation of the teleparallel geometry, where we understand the latter as also including the chosen orientation and time orientation as integral constituents, which we collectively denote by \(\epsilon\). This becomes even more clear by realizing that once the orientation and time orientation are fixed, the components of the metric \(g_{\mu\nu}\) can be used as local fiber coordinates on \(C\) (whose fibers are diffeomorphic to the 10-dimensional coset space \(G/H\)). Note, however, that these are not global coordinates, since the metric determines not a single point on each fiber of \(C\), but four points.

\item
From the discussion above we have learned that sections of \(C\) are in one-to-one correspondence of principal bundle reductions, which lead to a principal bundle \(P = \mathrm{SO}_0(M, g, \epsilon)\) with structure group \(\mathrm{SO}_0(1,3)\). From our assumption that \(M\) allows for global spinor fields, and therefore admits a spin structure, it follows that \(P\) must have a globally defined orthonormal frame, i.e., a global section \(e: M \to P\)~\cite{Geroch:1968zm}. Denoting the dual coframe by \(\theta\), and defining the spin connection as the pullback \(\omega = e^*\Omega \in \Omega^1(M, \mathfrak{h})\), one thus obtains a teleparallel geometry in the sense of the covariant formulation of teleparallel gravity. It encodes the metric, orientation, time orientation and the teleparallel connection. The freedom to choose the local section \(e\) corresponds to the local Lorentz invariance of this formulation.

\item
Using the procedure given in section~\ref{ssec:weitz}, the tetrad \(\theta\) can locally be transformed into the Weitzenböck gauge. This transformation corresponds to the choice of a \emph{local} section \(\wb{e}: U \to P\) on a simply connected neighborhood \(U \subset M\) which is horizontal with respect to the connection \(\Omega\), whence \(\wb{\omega} = \wb{e}^*\Omega = 0\).
\end{enumerate}

To summarize, we see that the conventional formulations of the teleparallel geometry are simply different possibilities to assign (in general only local) coordinates to the more general, abstract geometric structure described in the beginning of this section, and are thus locally equivalent (although, in general, not globally).

\subsection{Refinement using spin structure}\label{ssec:spinor}
One of the main physical arguments we used to conclude on the existence of a global tetrad on a physically meaningful spacetime was the necessity to describe fermions by spinor fields, which requires the existence of a spin structure~\cite{Rajan:2016qiy}. In fact, the latter can be obtained from the former by the following procedure. Let \(e: M \to P\) be a section of the (trivial) oriented, time-oriented orthonormal frame bundle, and \(Q = M \times \mathrm{SL}(2, \mathbb{C})\). Using the covering map \(\sigma: \mathrm{SL}(2, \mathbb{C}) \to \mathrm{SO}_0(1,3)\), one can then define a map \(\varphi: Q \to P\) by \(\varphi(x, z) = e(x) \cdot \sigma(z)\). Clearly, this is a two-fold covering map, since \(\sigma(-z) = \sigma(z)\), and hence a spin structure. Of course one may pose the question whether every spin structure can be described using such a trivial bundle \(Q\), or whether more general cases exist. It turns out that the former is true: whenever there exists a spin structure on a spacetime \(M\), it necessarily admits a global section \(s: M \to Q\), and so it must be trivial~\cite{Geroch:1968zm}. Choosing this to be \(s(x) = (x, \mathds{1}_2)\), one has \(e = \varphi \circ s\), and so one recovers the tetrad as well.

Naturally, the question arises whether any two tetrads \(e, e'\) give rise to the same spin structure. Naively one may think that this is the case, since the total space \(Q\) and its projection to \(M\) are independent of the choice of the tetrad. However, note that also the covering map \(\varphi: Q \to P\) is a part of the definition of a spin structure, and two spin structures \(\varphi: Q \to P\) and \(\varphi': Q \to P\) can be regarded as equivalent only if there exists a principal bundle isomorphism \(\mu: Q \to Q'\) such that \(\varphi = \varphi' \circ \mu\). The latter, however, does not hold in general, unless \(M\) is simply connected. This can be seen as follows:

Consider two (global) tetrads \(e, e': M \to P\), and assume for simplicity \(e(x) = e'(x)\) for some point \(x \in M\). Then there exists a local Lorentz transformation \(\Lambda: M \to \mathrm{SO}_0(1,3)\) such that \(e = e' \cdot \Lambda\) with \(\Lambda(x) = \mathds{1}_4\). If \(M\) is not simply connected, we can choose these tetrads such that there exists a path \(\gamma: [0, 1] \to M\) with \(\gamma(0) = \gamma(1) = x\) and \(\Lambda \circ \gamma: [0, 1] \to \mathrm{SO}_0(1,3)\) is not contractible. (This is possible since \(\mathrm{SO}_0(1,3)\) is not simply connected.) Let \(\varphi: Q \to P\) and \(\varphi': Q \to P\) be the spin structures constructed by the procedure above from \(e\) and \(e'\), and \(s: M \to Q, x \mapsto (x, \mathds{1}_2)\). By construction, we have \(e = \varphi \circ s\) and \(e' = \varphi' \circ s\). Now assume that there exists a principal bundle isomorphism \(\mu: Q \to Q'\) such that \(\varphi = \varphi' \circ \mu\). Since \(\varphi\) and \(\varphi'\) are principal bundle homomorphisms by definition of a spin structure, and thus fiber preserving and equivariant, one must have
\begin{equation}
e'(y)\Lambda(y) = e(y) = \varphi(s(y)) = \varphi'(\mu(s(y)))\,,
\end{equation}
and so \(\mu(s(y)) = s(y)\Sigma(y)\) with \(\Sigma: M \to \mathrm{SL}(2, \mathbb{C})\) such that \(\Lambda = \sigma \circ \Sigma\). In particular, for \(y = x\) we have \(\sigma(\Sigma(x)) = \mathds{1}_4\), and thus \(\Sigma(x) \in \{\pm\mathds{1}_2\}\). Now consider \(\Sigma \circ \gamma: [0, 1] \to \mathrm{SL}(2, \mathbb{C})\). Since \(\sigma \circ \Sigma \circ \gamma\) is not contractible, it must wind around \(\mathrm{SO}_0(1,3)\), and so one must have \(\Sigma(\gamma(0)) = -\Sigma(\gamma(1))\). But \(\gamma(0) = \gamma(1) = x\), leading to a contradiction. Hence, no such principal bundle isomorphism \(\mu\) can exist.

This answers our question for the equivalence of the spin structures defined by two tetrads \(e, e'\): they are equivalent if and only if there exists a local spin transformation \(\Sigma: M \to \mathrm{SL}(2, \mathbb{C})\) such that \(e = e' \circ (\sigma \circ \Sigma)\). To see that equivalence indeed follows from the latter, set \(\mu(x, u) = (x, u\Sigma(x))\), and observe that
\begin{equation}
\varphi'(\mu(s(x))) = \varphi'(x, \Sigma(x)) = \varphi'(x, \mathds{1}_2)\sigma(\Sigma(x)) = e'(x)\sigma(\Sigma(x)) = e(x) = \varphi(s(x))\,,
\end{equation}
and hence \(\varphi = \varphi' \circ \mu\) by equivariance.

We finally return to the question which fundamental geometric object is most appropriate as describing a teleparallel geometry. Demanding that this object uniquely and unambiguously defines the geometric background for physical observations and the dynamics of physical fields, including spinor fields, we now conclude that it must define not only a metric, orientation and time orientation, but also a spin structure. From our findings above, we see that this object is an equivalence class of global tetrads, or equivalently global coframes \(\theta^a{}_{\mu}\), where equivalence is defined up to a local spin transformation \(\Sigma: M \to \mathrm{SL}(2, \mathbb{C})\). Further demanding that it fully describes the gravitational interaction of teleparallel gravity, this is further complemented by a spin connection \(\omega^a{}_{b\mu}\), where equivalence is now regarded with respect to simultaneous local Lorentz (spin) transformations of both the tetrad and the spin connection.

\section{Conclusion}\label{sec:conclusion}
In this article we discussed the different formulations of teleparallel gravity from a geometric perspective, and studied the question which geometric object is most appropriate for describing a teleparallel geometry. Based on physical arguments, such as the demand to describe physical observations using clocks and light propagation, causality, parity-violating interactions and spinor fields, besides providing the dynamical fields of teleparallel gravity, we concluded that the most concise description of a teleparallel geometry is given by an equivalence class \([\theta^a{}_{\mu}, \omega^a{}_{b\mu}]\) of a (globally defined) tetrad and spin connection, where equivalence is defined up to a local spin transformation \(\Sigma: M \to \mathrm{SL}(2, \mathbb{C})\). We have shown that such an equivalence class uniquely and unambiguously defines the geometric background for the aforementioned physical phenomena (a Lorentzian metric, orientation, time orientation and spin structure), as well as a flat, metric-compatible affine connection, whose torsion mediates the gravitational interaction in teleparallel gravity.

Further, we have compared the conventional formulations of teleparallel gravity theories to our geometric approach. We showed that the covariant formulation in terms of a tetrad and spin connection most closely resembles our result, provided that the local Lorentz gauge freedom is further restricted in order to preserve also the spin structure. We also presented a constructive transition from the covariant to the pure tetrad formulation, and showed that this is possible only locally, so that also these different formulations can only locally be regarded equivalent, while globally the covariant formulation is more general. We also discussed the Palatini formulation, and showed that it needs to be supplemented with an orientation, time orientation and spin structure in order to resemble our geometric formulation.

Finally, we have discussed the notions of local Lorentz invariance in the context of observational tests of gravity and of teleparallel gravity theories. We have used the aforementioned geometric interpretation, as well as the Palatini formulation of teleparallel gravity, in order to clarify that a breaking of the latter, which is a purely internal symmetry in teleparallel gravity and broken in the transition from TEGR to modified theories, does not imply a breaking of the former, which is strictly constrained by observations.

\section*{Acknowledgments}
The author gratefully acknowledges the full support by the Estonian Research Council through the Personal Research Funding project PRG356, as well as the European Regional Development Fund through the Center of Excellence TK133 ``The Dark Side of the Universe''.


\begin{thebibliography}{0}
%\bibitem{BeltranJimenez:2019tjy}
%J.~B.~Jiménez, L.~Heisenberg and T.~S.~Koivisto,
%The Geometrical Trinity of Gravity,
%Universe \textbf{5} (2019) no.7, 173
%[arXiv:1903.06830 [hep-th]].

\bibitem{Einstein:1928}
A.~Einstein,
Riemann-Geometrie mit Aufrechterhaltung des Begriffes des Fernparallelismus
Sitzber. Preuss. Akad. Wiss. \textbf{17} (1928) 217.

\bibitem{Aldrovandi:2013wha}
R.~Aldrovandi and J.~G.~Pereira,
Teleparallel Gravity : An Introduction,
Fundam.\ Theor.\ Phys. \textbf{173}, Springer, Dordrecht, 2013.

\bibitem{Bahamonde:2021gfp}
S.~Bahamonde, K.~F.~Dialektopoulos, C.~Escamilla-Rivera, G.~Farrugia, V.~Gakis, M.~Hendry, M.~Hohmann, J.~L.~Said, J.~Mifsud and E.~Di Valentino,
Teleparallel Gravity: From Theory to Cosmology,
[arXiv:2106.13793 [gr-qc]].

\bibitem{Cai:2015emx}
Y.~F.~Cai, S.~Capozziello, M.~De Laurentis and E.~N.~Saridakis,
f(T) teleparallel gravity and cosmology,
Rept. Prog. Phys. \textbf{79} (2016) no.10, 106901
[arXiv:1511.07586 [gr-qc]].

\bibitem{CANTATA:2021ktz}
E.~N.~Saridakis \textit{et al.} [CANTATA],
Modified Gravity and Cosmology: An Update by the CANTATA Network,
Springer, Cham, 2021
[arXiv:2105.12582 [gr-qc]].

\bibitem{Li:2010cg}
B.~Li, T.~P.~Sotiriou and J.~D.~Barrow,
$f(T)$ gravity and local Lorentz invariance,
Phys. Rev. D \textbf{83} (2011) 064035
[arXiv:1010.1041 [gr-qc]].

\bibitem{Sotiriou:2010mv}
T.~P.~Sotiriou, B.~Li and J.~D.~Barrow,
Generalizations of teleparallel gravity and local Lorentz symmetry,
Phys. Rev. D \textbf{83} (2011) 104030
[arXiv:1012.4039 [gr-qc]].

\bibitem{Golovnev:2021omn}
A.~Golovnev and M.~J.~Guzman,
Lorentz symmetries and primary constraints in covariant teleparallel gravity,
Phys. Rev. D \textbf{104} (2021) no.12, 124074
[arXiv:2110.11273 [gr-qc]].

\bibitem{Maluf:2013gaa}
J.~W.~Maluf,
The teleparallel equivalent of general relativity,
Annalen Phys. \textbf{525} (2013) 339
[arXiv:1303.3897 [gr-qc]].

\bibitem{Ferraro:2014owa}
R.~Ferraro and F.~Fiorini,
Remnant group of local Lorentz transformations in $\mathcal{f}(T)$ theories,
Phys. Rev. D \textbf{91} (2015) no.6, 064019
[arXiv:1412.3424 [gr-qc]].

\bibitem{Chen:2014qtl}
P.~Chen, K.~Izumi, J.~M.~Nester and Y.~C.~Ong,
Remnant Symmetry, Propagation and Evolution in $f(T)$ Gravity,
Phys. Rev. D \textbf{91} (2015) no.6, 064003
[arXiv:1412.8383 [gr-qc]].

\bibitem{Maluf:2018coz}
J.~W.~Maluf, S.~C.~Ulhoa, J.~F.~da Rocha-Neto and F.~L.~Carneiro,
Difficulties of Teleparallel Theories of Gravity with Local Lorentz Symmetry,
Class. Quant. Grav. \textbf{37} (2020) no.6, 067003
[arXiv:1811.06876 [gr-qc]].

\bibitem{Guzman:2019ozl}
M.~J.~Guzm\'an and R.~Ferraro,
Degrees of freedom and local Lorentz invariance in $f(T)$ gravity,
[arXiv:1903.06774 [gr-qc]].

\bibitem{Bejarano:2019fii}
C.~Bejarano, R.~Ferraro, F.~Fiorini and M.~J.~Guzm\'an,
Reflections on the covariance of modified teleparallel theories of gravity,
Universe \textbf{5} (2019) 158
[arXiv:1905.09913 [gr-qc]].

\bibitem{Ferraro:2020tqk}
R.~Ferraro and M.~J.~Guzm\'an,
Pseudoinvariance and the extra degree of freedom in $f(T)$ gravity,
Phys. Rev. D \textbf{101} (2020) no.8, 084017
[arXiv:2001.08137 [gr-qc]].

\bibitem{Will:2018bme}
C.~M.~Will,
Theory and Experiment in Gravitational Physics,
Cambridge University Press, Cambridge, 2018.

\bibitem{Will:2014kxa}
C.~M.~Will,
The Confrontation between General Relativity and Experiment,
Living Rev. Rel. \textbf{17} (2014) 4
[arXiv:1403.7377 [gr-qc]].

\bibitem{Pfeifer:2014yua}
C.~Pfeifer,
Radar orthogonality and radar length in Finsler and metric spacetime geometry,
Phys. Rev. D \textbf{90} (2014) no.6, 064052
[arXiv:1408.5306 [gr-qc]].

\bibitem{Pfeifer:2011tk}
C.~Pfeifer and M.~N.~R.~Wohlfarth,
Causal structure and electrodynamics on Finsler spacetimes,
Phys. Rev. D \textbf{84} (2011) 044039
[arXiv:1104.1079 [gr-qc]].

\bibitem{Hohmann:2021zbt}
M.~Hohmann, C.~Pfeifer and N.~Voicu,
Finsler-based field theory -- a mathematical foundation,
[arXiv:2106.14965 [math-ph]].

\bibitem{Baum:2014}
H.~Baum,
Eichfeldtheorie,
Springer, Berlin \& Heidelberg, 2014.

\bibitem{Ualikhanova:2019ygl}
U.~Ualikhanova and M.~Hohmann,
Parametrized post-Newtonian limit of general teleparallel gravity theories,
Phys. Rev. D \textbf{100} (2019) no.10, 104011
[arXiv:1907.08178 [gr-qc]].

\bibitem{Emtsova:2019qsl}
E.~D.~Emtsova and M.~Hohmann,
Post-Newtonian limit of scalar-torsion theories of gravity as analogue to scalar-curvature theories,
Phys. Rev. D \textbf{101} (2020) no.2, 024017
[arXiv:1909.09355 [gr-qc]].

\bibitem{Flathmann:2019khc}
K.~Flathmann and M.~Hohmann,
Post-Newtonian Limit of Generalized Scalar-Torsion Theories of Gravity,
Phys. Rev. D \textbf{101} (2020) no.2, 024005
[arXiv:1910.01023 [gr-qc]].

\bibitem{Bahamonde:2020cfv}
S.~Bahamonde, K.~F.~Dialektopoulos, M.~Hohmann and J.~Levi Said,
Post-Newtonian limit of Teleparallel Horndeski gravity,
Class. Quant. Grav. \textbf{38} (2020) no.2, 025006
[arXiv:2003.11554 [gr-qc]].

\bibitem{Ferraro:2011us}
R.~Ferraro and F.~Fiorini,
Non trivial frames for $f(T)$ theories of gravity and beyond,
Phys.\ Lett.\ B \textbf{702} (2011) 75
[arXiv:1103.0824 [gr-qc]].

\bibitem{Tamanini:2012hg}
N.~Tamanini and C.~G.~Boehmer,
Good and bad tetrads in $f(T)$ gravity,
Phys.\ Rev.\ D \textbf{86} (2012) 044009
[arXiv:1204.4593 [gr-qc]].

\bibitem{Krssak:2015oua}
M.~Krššák and E.~N.~Saridakis,
The covariant formulation of $f(T)$ gravity,
Class.\ Quant.\ Grav. \textbf{33} (2016) 115009
[arXiv:1510.08432 [gr-qc]].

\bibitem{Golovnev:2017dox}
A.~Golovnev, T.~Koivisto and M.~Sandstad,
On the covariance of teleparallel gravity theories,
Class. Quant. Grav. \textbf{34} (2017) no.14, 145013
[arXiv:1701.06271 [gr-qc]].

\bibitem{Krssak:2018ywd}
M.~Krššák, R.~J.~van den Hoogen, J.~G.~Pereira, C.~G.~Böhmer and A.~A.~Coley,
Teleparallel theories of gravity: illuminating a fully invariant approach,
Class.\ Quant.\ Grav. \textbf{36} (2019) 183001
[arXiv:1810.12932 [gr-qc]].

\bibitem{Hohmann:2018rwf}
M.~Hohmann, L.~Järv and U.~Ualikhanova,
Covariant formulation of scalar-torsion gravity,
Phys.\ Rev.\ D \textbf{97} (2018) 104011
[arXiv:1801.05786 [gr-qc]].

\bibitem{Rajan:2016qiy}
D.~Rajan and M.~Visser,
`Global properties of physically interesting Lorentzian spacetimes,
Int. J. Mod. Phys. D \textbf{25} (2016) no.14, 1650106
[arXiv:1601.03355 [gr-qc]].

\bibitem{Geroch:1968zm}
R.~P.~Geroch,
Spinor structure of space-times in general relativity. I,
J. Math. Phys. \textbf{9} (1968) 1739.

\bibitem{BeltranJimenez:2018vdo}
J.~Beltr\'an Jim\'enez, L.~Heisenberg and T.~S.~Koivisto,
Teleparallel Palatini theories,
JCAP \textbf{08} (2018), 039
[arXiv:1803.10185 [gr-qc]].

\bibitem{Hohmann:2021fpr}
M.~Hohmann,
Variational Principles in Teleparallel Gravity Theories,
Universe \textbf{7} (2021) no.5, 114
[arXiv:2104.00536 [gr-qc]].

\bibitem{Obukhov:2002tm}
Y.~N.~Obukhov and J.~G.~Pereira,
Metric affine approach to teleparallel gravity,
Phys. Rev. D \textbf{67} (2003) 044016
[arXiv:gr-qc/0212080 [gr-qc]].

\bibitem{Maluf:2003fs}
J.~W.~Maluf,
Dirac spinor fields in the teleparallel gravity: Comment on `Metric affine approach to teleparallel gravity',
Phys. Rev. D \textbf{67} (2003) 108501
[arXiv:gr-qc/0304005 [gr-qc]].

\bibitem{Mosna:2003rx}
R.~A.~Mosna and J.~G.~Pereira,
Some remarks on the coupling prescription of teleparallel gravity,
Gen. Rel. Grav. \textbf{36} (2004) 2525
[arXiv:gr-qc/0312093 [gr-qc]].

\bibitem{Mielke:2004gg}
E.~W.~Mielke,
Consistent coupling to Dirac fields in teleparallelism: Comment on `Metric-affine approach to teleparallel gravity',
Phys. Rev. D \textbf{69} (2004) 128501.

\bibitem{Obukhov:2004hv}
Y.~N.~Obukhov and J.~G.~Pereira,
Lessons of spin and torsion: Reply to `Consistent coupling to Dirac fields in teleparallelism',
Phys. Rev. D \textbf{69} (2004) 128502
[arXiv:gr-qc/0406015 [gr-qc]].

\bibitem{BeltranJimenez:2020sih}
J.~Beltr\'an Jim\'enez, L.~Heisenberg and T.~Koivisto,
The coupling of matter and spacetime geometry,
Class. Quant. Grav. \textbf{37} (2020) no.19, 195013
[arXiv:2004.04606 [hep-th]].

\bibitem{Golovnev:2020zpv}
A.~Golovnev and M.~J.~Guzm\'an,
Foundational issues in f(T) gravity theory,
Int. J. Geom. Meth. Mod. Phys. \textbf{18} (2021) no.supp01, 2140007
[arXiv:2012.14408 [gr-qc]].

\bibitem{BeltranJimenez:2019nns}
J.~Beltr\'an Jim\'enez and K.~F.~Dialektopoulos,
Non-Linear Obstructions for Consistent New General Relativity,
JCAP \textbf{01} (2020) 018
[arXiv:1907.10038 [gr-qc]].

\bibitem{Jimenez:2020ofm}
J.~B.~Jim\'enez, A.~Golovnev, T.~Koivisto and H.~Veerm\"ae,
Minkowski space in $f(T)$ gravity,
Phys. Rev. D \textbf{103} (2021) no.2, 024054
[arXiv:2004.07536 [gr-qc]].

\bibitem{Golovnev:2020nln}
A.~Golovnev and M.~J.~Guzman,
Nontrivial Minkowski backgrounds in $f(T)$ gravity,
Phys. Rev. D \textbf{103} (2021) no.4, 044009
[arXiv:2012.00696 [gr-qc]].

\bibitem{Kopczynski:1982}
W.~Kopczynski,
Problems with metric-teleparallel theories of gravitation,
J. Phys. A: Math. Gen. \textbf{15} (1982) 493.

\bibitem{Nester:1988}
J.~M.~Nester,
Is there really a problem with the teleparallel theory?,
Class. Quant. Grav. \textbf{5} (1988) 1003.

\bibitem{Ong:2013qja}
Y.~C.~Ong, K.~Izumi, J.~M.~Nester and P.~Chen,
Problems with Propagation and Time Evolution in $f(T)$ Gravity,
Phys. Rev. D \textbf{88} (2013) 024019
[arXiv:1303.0993 [gr-qc]].

\bibitem{Izumi:2013dca}
K.~Izumi, J.~A.~Gu and Y.~C.~Ong,
Acausality and Nonunique Evolution in Generalized Teleparallel Gravity,
Phys. Rev. D \textbf{89} (2014) no.8, 084025
[arXiv:1309.6461 [gr-qc]].

\bibitem{Golovnev:2018wbh}
A.~Golovnev and T.~Koivisto,
Cosmological perturbations in modified teleparallel gravity models,
JCAP \textbf{11} (2018) 012
[arXiv:1808.05565 [gr-qc]].

\bibitem{Golovnev:2020aon}
A.~Golovnev,
Perturbations in $f(\mathbb{T})$ cosmology and the spin connection,
JCAP \textbf{2004} (2020) 014
[arXiv:2001.10015 [gr-qc]].
\end{thebibliography}
\end{document}